\documentclass[graybox]{svmult} 
\usepackage{helvet}         
\usepackage{courier}        
\usepackage{type1cm}        
\usepackage{makeidx}         
\usepackage{graphicx}        
\usepackage{multicol}        
\usepackage[bottom]{footmisc}

\usepackage{ifpdf}
\usepackage{ae}
\usepackage[T1]{fontenc}
\usepackage[ansinew]{inputenc}
\usepackage{mathrsfs}
\usepackage{amsmath}
\usepackage{amssymb}
\usepackage{dsfont}
\pdfoutput=1
\usepackage{amsmath,amssymb,amsfonts,a4wide,graphicx,bm,times,psfrag,wrapfig,sidecap}
\usepackage{cite}
\usepackage[colorlinks=true,linkcolor=black, citecolor=black,
urlcolor=black]{hyperref}
\numberwithin{equation}{section}
\makeatletter \let\old@startsection=\@startsection
\renewcommand{\@startsection}[6]
{\old@startsection{#1}{#2}{#3}{#4}{#5}{#6\mathversion{bold}}}
\makeatother
\def\O{\Omega}
\def\Det{ \text{Det}}

\newcommand\re[1]{({\ref{#1}})}
\def\be{\begin{eqnarray}  }
    \def\ee{\end{eqnarray}}

    \def\ID{{\mathbb{D}}}

\def\ii{i \varepsilon}
    \def\no{\nonumber}
    \def\la{\label}

\def\({\left(} \def\){\right)}
\def\<{\left\langle\,}
\def\>{\, \right\rangle}
\def\[{\left[}
 \def\]{\right]}
\def\tr{{\rm   tr} }
    \def\hf{ {\textstyle{1\over 2}} }
       
    \def\CA{{\cal A}}

\def\CO{{ \mathcal{ O} }}

    \def\CK{{ \mathcal{ K} }}

     \def\L{\Lambda}
     \def\CC{ {\mathcal C}}
       
       \def\CN{{ \cal  N}}

      \def\CV{{ \cal  V}}

 \def\CY{{\cal Y}}
  \def\p{\partial}
  \def\a{\alpha}
 \def\b{\beta}

  \def\ve{\varepsilon}

  \def\vp{\varphi}
  \def\th{\theta}

\newcommand{\caA}{{\mathscr A}}

\def\ket{ | 0 \rangle}
\def\bra{ \langle 0 | }
\def\zz{ { \{ \bf{\sigma} \} }}
  \def\Vii{\CV  }
   \def\CQ{{\cal Q}}

\newcommand\encadremath[1]{\vbox{\hrule\hbox{\vrule\kern8pt
\vbox{\kern8pt \hbox{$\displaystyle #1$}\kern8pt}
\kern8pt\vrule}\hrule}} \def\enca#1{\vbox{\hrule\hbox{
\vrule\kern8pt\vbox{\kern8pt \hbox{$\displaystyle #1$} \kern8pt}
\kern8pt\vrule}\hrule}}

  \usepackage{bm}
\def\O{\Omega}

\def\ee{\end{eqnarray}}  

    \def\no{\nonumber} \def\la{\label} 
\def\({\left(} \def\){\right)}
\def\<{\left\langle\,} \def\>{\,
\right\rangle}
\def\[{\left[} \def\]{\right]} \def\tr{{\rm tr} }
\def\hf{ {\textstyle{1\over 2}} } 
\def\CA{{\cal A}}   \def\CO{{
\mathcal{ O} }}  
\def\CK{{ \mathcal{ K} }} 
\def\L{\Lambda}
     \def\CC{ {\mathcal C}}  \def\CN{{ \cal N}}
       
       \def\CY{{\cal Y}}
     \def\p{\partial} \def\a{\alpha} \def\b{\beta} 
        \def\vp{\varphi}
     \def\th{\theta}  
     
   \def\Li{ \text{Li}_2}

\def\zz{ { { \bf  z} }}

\def\uu{ { {\bf u} }}

\def\vv{ { \bf v}}
\def\ww{ { \bf w }}
\def\thth{ { \bm {\th } }}
\def\vv { { \bf v  }}

  \def\k{\kappa}

   \def\k{\kappa}

 \def\CQ{{\cal Q}}

\begin{document}

\title*{Semi-classical  scalar products in the generalised $SU(2)$ model}

 \author{Ivan Kostov}

\institute{Ivan Kostov \at Institut de Physique Th\'eorique,
CNRS-URA 2306, C.E.A.-Saclay, F-91191 Gif-sur-Yvette, France\\
{\it Associate member of the Institute for Nuclear Research and
Nuclear Energy, Bulgarian Academy of Sciences, 72 Tsarigradsko
Chauss\'ee, 1784 Sofia, Bulgaria}}


\maketitle

   \abstract{In these notes we  review the field-theoretical approach to the computation of
 the  scalar product  of multi-magnon states in the  Sutherland limit
 where the magnon rapidities  condense into one or several macroscopic arrays.
 We formulate a systematic procedure for
computing the  $1/M$ expansion
of the on-shell/off-shell scalar product of  $M$-magnon states
in the generalised integrable model with $SU(2)$-invariant rational $R$-matrix.
 The coefficients of the expansion
are obtained as multiple contour integrals in the rapidity plane.}

%

\vskip 2cm

\centerline{\it \footnotesize 
Based on the talk by the author delivered at the  }
  \centerline{\it\footnotesize 
X. International Workshop: ``Lie Theory and Its Applications in Physics'' (LT-10),
Varna, Bulgaria, 17-23 June 2013
}

\newpage
\setcounter{footnote}{0}

\section{Introduction}

In many cases the calculation of form factors and correlation
functions within quantum integrable models solvable by the Bethe
Ansatz reduces to the calculation of scalar products of Bethe vectors.
The best studied case is that of the models based on the
$SU(2)$-invariant $R$-matrix.  A determinant formula for the the
norm-squared of an on-shell state has been conjectured by Gaudin
\cite{Gaudin-livre}, and then proved by Korepin in
\cite{korepin-DWBC}.  Sum formulas for the scalar product between two
generic Bethe states were obtained by by Izergin and Korepin
\cite{korepin-DWBC,2009arXiv0911.1881K,Izergin:1984aa}.  Furthermore, the scalar
product between an on-shell and off-shell Bethe vector was expressed
in determinant form by Slavnov \cite{NSlavnov1}.  This representation
proved to be very useful in the computation of correlation functions
of the XXX and XXZ models \cite{2005hep.th....5006K}.  Although the
Slavnov determinant formula is, by all evidence, not generalisable for
higher rank groups, compact and potentially useful expressions of the
scalar products as multiple contour integrals of (products of)
determinants were proposed in \cite{2005TMP...145.1373P,
2009AnHP...10..513F,2010SIGMA...6..094B,2013arXiv1306.0552W}.

 The above-mentioned sum and determinant formulas
 are  efficient for states compsed if few magnons.  In order to evaluate
 scalar products of multi-magnon states, new semi-classical methods
 specific for the problem need to be developed.

 Of particular interest is the evaluation of the scalar product
 of Bethe wave functions  describing the lowest excitations above the
ferromagnetic vacuum composed of given (large) number of magnons.
The magnon rapidities for such excitations organise themselves in a small
number of macroscopically large bound complexes
 \cite{PhysRevLett.74.816, PhysRevLett.85.2813}.
It is common to refer this limit as a thermodynamical, or semi-classical,
or Sutherland limit.
  In the last years
the thermodynamical limit attracted much attention in the context of
the integrability in AdS/CFT \cite{ReviewIGST}, where it describes
``heavy'' operators in the $\CN=4$ supersymmetric Yang-Mills theory
(SYM), dual to classical strings embedded in the curved $AdS_5\times
S^5$ space-time \cite{Beisert:2003xu,Kazakov:2004qf}.  It has been
realised that the computation of some 3-point functions of such heavy
operators boils down to the computation of the scalar product of the
corresponding Bethe wave functions in the thermodynamical limit
\cite{EGSV,Foda:3ptdeterminant,3pf-prl,SL,2014arXiv1401.0384J}.

 In this notes, based largely on the results obtained in
\cite{3pf-prl,SL,sz,Eldad-Ivan}, we  review
 the field-theoretical approach developed by E. Bettelheim and the author
\cite{Eldad-Ivan}, which leads to a systematical semi-classical expansion of
the on-shell/off-shell scalar product. The field-theoretical representation
is not sensitive to the particular representation of the monodromy matrix and
we put it in the context of the generalised integrable model with
$SU(2)$ invariant rational $R$-matrix.

  The text is organised as follows.  In Section \ref{ABA} we remind
  the basic facts and conventions concerning the Algebraic Bethe
  Ansatz for rational $SU(2)$-invariant $R$-matrix.  In Section
  \ref{DetScl} we give an alternative determinant representation of
  the on-shell/off-shell scalar product of two $M$-magnon Bethe
  vectors in spin chains with rational $SU(2)$-invariant $R$-matrix.
  This representation, which has the form of an $2M\times 2M$
  determinant, possesses an unexpected symmetry: it is invariant under
  the group $S_{2M}$ of the permutations of the {\it union} of the
  magnon rapidities of the left and the right states, while the
  Korepin sum formulas and the Slavnov determinant have a smaller
  $S_M\times S_M$ symmetry.  We refer to the  symmetric
  expression in question as $\caA$-functional to underline the relation with a
  similar quantity, previously studied in the papers \cite{EGSV, GSV}
  and denoted there by the same letter.  In the generalised $SU(2)$-invariant
  integrable model the $\caA$-functional depends on the ratio of the
  eigenvalues of the diagonal elements of the monodromy matrix on the
  pseudo-vacuum, considered as a free functional variable.  In Section
  \ref{FieldTheory} we write the $\caA$-functional as an expectation
  value in the Fock space of free chiral fermions.  The fermionic
  representation implies that the $\caA$-functional is a KP
  $\tau$-function, but we do not use this fact explicitly.  By
  two-dimensional bosonization we obtain a formulation of the
  $\caA$-functional in terms of a chiral bosonic field with
  exponential interaction.  The bosonic field describes a Coulomb gas
  of dipole charges.  The thermodynamical limit $M\gg1$ is described
  by an effective $(0+1)$-dimensional field theory, obtained by
  integrating  the fast-scale modes of the original bosonic field.  In
  terms of the dipole gas the effective theory contains composite
  particles representing bound states of any number of dipoles.  The
  Feynman diagram technique for the effective field theory for the
  slow-scale modes is expected to give the perturbative $1/M$
  expansion of the scalar product.  We evaluate explicitly the first
  two terms of this expansion.  The leading term reproduces the known
  expression as a contour integral of a dilogarithm, obtained by
  different methods in \cite{GSV} and \cite{3pf-prl,SL}, while the
  subleading term, given by a double contour integral, is a new result
  reported recently in \cite{Eldad-Ivan}.

\section{Algebraic Bethe Ansatz for integrable models with $su(2)$
$R$-matrix}
\la{ABA}

 We remind some facts about the ABA for the $su(2)$-type models and
 introduce our notations.  The monodromy matrix $M(u)$ is a $2\times
 2$ matrix \cite{Takhtajan:1979iv,Faddeev:1979gh} \be
\label{monodABCD}
M (u)=\left(\begin{array}{cc}A(u) & B(u) \\ C(u) & D(u)\end{array}\right).
\ee
The matrix elements $A, B, C, D$ are operators in the Hilbert space of
the model and depend on the complex spectral parameter $u$ called
rapidity.  The monodromy matrix obeys the $RTT$-relation (Yang-Baxter
equation)
\begin{eqnarray}
\la{RTT} R(u-v) (M (u)\otimes I)(I\otimes M(v)) =(I\otimes M(u)) (M
(v)\otimes I)R(u-v).
\end{eqnarray}
 Here $I$ denotes the $2\times 2$ identity matrix and the $4\times 4$
 matrix $R(u)$ is the $SU(2)$ rational $R$-matrix whose entries are
 c-numbers.  The latter is given, up to a numerical factor, by
\begin{equation}
\la{defRmatrix}
R_{\a\b}(u)=u\, I_{\a\b}+ {\ii} \, P_{\a\b} ,
\end{equation}
with  the operator $P_{\a\b}$ acting as  a permutation of the spins  in
the spaces $\a$ and $\b$.
In the standard normalization $\ve=1$.

The $RTT$ relation determines the algebra of the monodromy matrix
elements, which is the same for all $su(2)$-type models.  In
particular, $[B(u), B(v)] = [C(u),C(v)]= 0$ for all $u$ and $v$.

The trace $T= A+D$ of the monodromy matrix is called transfer matrix.
Sometimes it is useful to introduce a twist parameter $\k$ (see, for
example, \cite{nikitaslavnov}).  The twist preserves the integrability:
the  twisted transfer matrix
\begin{equation}
T(u)=\tr \[ ( ^{1\ 0}_{0\ \k})M (u)\] =A(u)+\k\, D(u)
\end{equation}
 satisfies $[T(u), T(v)]=0$ for all $u$ and $v$.

To define a quantum-mechanical system completely, one must determine
the action of the elements of the monodromy matrix in the Hilbert
space.  In the framework of the ABA the Hilbert space is constructed
as a Fock space associated with a cyclic vector $|\Omega\rangle$,
called pseudovacuum, which is an eigenvector of the operators $A$ and
$D$ and is annihilated by the operator $C$:
\be
\la{defadket}
A(u)|\Omega\rangle= a(u) |\Omega\rangle, \
D(u)|\Omega\rangle=d(u)|\Omega\rangle,\
C(u)|\Omega\rangle=0.
\ee
The dual pseudo-vacuum  satisfies the relations
\be
\la{defadbra}
 \langle   \Omega | A(u) = a(u)  \langle  \Omega | ,\
 \langle  \Omega | D(u) =d(u) \langle  \Omega| ,\
  \langle  \Omega | B(u) =0.
\ee
Here $a(u)$ and $d(u)$ are are complex-valued functions whose explicit
form depends on the choice of the representation of the algebra
\re{RTT}.  We will not need the specific form of these functions,
except for some mild analyticity requirements.  In other words, we
will consider the generalized $SU(2)$ model in the sense of
\cite{korepin-DWBC}, in which the functions $a(u)$ and $d(u)$ are
considered as free functional parameters.

The vectors obtained from the pseudo-vacuum $|\Omega \rangle $ by
acting with the `raising operators' $B(u)$,
\begin{eqnarray}
|\uu \rangle =B(u_1)\ldots B(u_M)|\Omega\rangle \; ,
\quad \uu=\{ u_1,\dots, u_M\}
\end{eqnarray}
are called {\it Bethe states}.  Since the $B$-operators commute, the
state $|\uu\rangle$ is invariant under the permutations of the
elements of the set $\uu$.

The Bethe states that are eigenstates of the (twisted) transfer matrix are
called `on-shell'.  Their rapidities obey the Bethe Ansatz equations
\be
\label{eq:BAEin}
{a(u_j)\over d(u_j)}
+  \k\, {Q_\uu( u_j+\ii)\over Q_\uu(u_j-\ii)}=1 \qquad (j=1,\dots, M).
\ee
Here and in the following we will use the notation
\begin{align}
 \la{defBaxP} Q_{\uu}(v) = \prod_{i=1}^M (v-u_i), \quad \uu
 &=\{u_1,\ldots,u_M\}.  \qquad
\end{align}
The corresponding  eigenvalue of the transfer matrix $T(x)$
is
%
\begin{eqnarray}
t(v)=\frac{Q_\uu(v-\ii)}{Q_\uu(v)}+\k\,
\frac{d(v)}{a(v)}\frac{Q_\uu(v+\ii)}{Q_\uu(v)} \;.
\end{eqnarray}
%
%
If the rapidities $\uu $ are generic, the Bethe state
is called `off-shell'.

In the unitary  representations of the $RTT$-algebra, like the XXX$_{1/2}$ spin
chain, the on-shell states form a complete set in the Hilbert space.
The XXX spin chain of length $L$ can be deformed by introducing
inhomogeneities $\th_1, \dots, \th_L$ associated with the $L$ sites of
the spin chain.  The eigenvalues of the operators $A(v)$ and $D(v)$ on
the vacuum in the inhomogeneous XXX chain are given by
\begin{eqnarray}
\la{adinhom}
  a(v)=Q_\thth(v+\hf  \ii) \; , &  &
d(v) =
{Q_\thth(v- \hf \ii) },
\;
\end{eqnarray}
where the polynomial $\CQ _\thth(x)$ is defined as\footnote{This is a
particular case of the Drinfeld polynomial  $P_1(u)$ \cite{DRINFELDP}  when all spins
along the chain are equal to $1/2$.}
\be
Q_\thth (x)=
\prod_{l=1}^L (x - \th_l),
\quad
\thth=\{ \th_1,\dots, \th_L\}.
\ee

  Any Bethe state is completely characterised by its {\it
  pseudo-momentum}, known also under the name of {\it counting
  function} \cite{DeVega-CF}
   \be \la{pseudomomentum} 2 i p (v) = \log {Q_\uu (v+\ii)\over
   Q_\uu(v-\ii)}-\log {a(v)\over d(v)}+ \log \k .  \ee
   The Bethe equations \re{eq:BAEin}
   imply that
\be
p(u_j) = 2\pi n_j -\pi \quad
(j=1, \dots, M)
\ee
 where the integers $n_j$ are called mode numbers.
%

       \section{Determinant  formulas for the inner product }
   \la{DetScl}

In order to expand the states $|\vv\rangle $ with given a set of
rapidities in the basis of eigenvectors $|\uu\rangle $ of the
monodromy matrix,
\be
|\vv\rangle =\sum_{\uu \ {\rm on \ shell}} {\< \uu|\vv\> \over \< \uu|\uu\>} \ |\uu\rangle ,
\ee
  we need to compute the scalar product $\< \vv|\uu\> $ of an
  off-shell and an on-shell Bethe state.  The scalar product is
  related to the bilinear form
  \be \la{definprod} (\vv, \uu) = \langle \O | \prod_{j=1}^M C(v_j)\
  \prod_{j=1}^M B(u_j) |\O\rangle \ee
  by $ \(\uu,\vv \)= (-1)^M \< \uu^*|\vv\>$.  This follows from the
  complex Hermitian convention $B(u)^\dag = - C(u^*)$.  The inner
  product can be computed by commuting the $B$-operators to the left
  and the $A$-operators to the right according to the algebra
  \re{RTT}, and then applying the relations \re{defadket} and
  \re{defadbra}.  The resulting sum formula written down by Korepin
   \cite{korepin-DWBC} works
  well for small number of magnons but for larger $M$ becomes
  intractable.

An important observation was made by N. Slavnov \cite{NSlavnov1}, who
realised that when one of the two states is on-shell, the Korepin sum
formula gives the expansion of the determinant of a sum of two
$M\times M$ matrices.\footnote{This property is particular for the
$SU(2)$ model.  The the inner product in the $SU(n)$ model is a
determinant only for a restricted class of states \cite{Wheeler-SU3}.}
Although the Slavnov determinant formula does not give obvious
advantages for taking the thermodynamical limit, is was used to
elaborate alternative determinant formulas, which are better suited
for this task \cite{3pf-prl,SL,sz,Eldad-Ivan}.

Up to a trivial factor, the inner product depends on the functional
argument
\be f(v) \equiv \k\, {d(v)\over a(v)} \ee 
and on two sets of
rapidities, $\uu=\{u_1, \dots, u_M\}$ and $ \vv= \{ v_1, \dots,
v_M\}$.  Since the rapidities within each of the two sets are not
ordered, the inner product has symmetry $S_M\times S_M$, where $S_M$
is the group of permutations of $M$ elements.
It came then as a surprise that the inner product can be written  \cite{sz} \footnote{The case considered in \cite{sz} was that of the
periodic inhomogeneous XXX$_{1/2}$ spin chain of length $L$, but the
proof given there is trivially extended to the generalised $SU(2)$ model.  }  as
a restriction on the mass shell (for one of the two sets of rapidities)
of an expression completely symmetric with
respect of the permutations of the union $ \ww\equiv \{w_1, \dots ,
w_{2M}\} = \{ u_1, \dots , u_M, v_1,\dots, v_M\} $ of the rapidities
of the two states:
   \be \la{defSuv0} ( \vv \vert \uu) &\underset{\uu \to
    \text{on shell}}=&\prod_{j =1}^M a(v_j ) d(u_j
   )\ \caA_\ww [f]\, , \qquad \ww=\uu\cup\vv , \ee
where the functional $\caA_\ww[f]$
is given by the following $N\times
N$ determinant ($N=2M$)
 \be \la{detformulaA} \caA_\ww[f] &=& { \det_{j k} \( w_j ^{k-1} -
 f(w_j ) \, (w_j +\ii )^{k-1}\)/ \det_{j k} \( w_j ^{k-1}\) } \, .
 \ee

In the XXX$_{1/2}$ spin
chain, the r.h.s. of \re{defSuv0} is proportional to the inner product
of an off-shell Bethe state
$|\ww\rangle$ and a state obtained from the left vacuum by a global
$SU(2)$ rotation \cite{sz}. Such inner products can be given statistical interpretation as
a partial domain-wall partition function (pDWPF) \cite{FW3}.
 In this case the identity \re{defSuv0} can be explained  with  the global $su(2)$
symmetry \cite{sz}.

Another determinant formula, which is particularly useful for taking
the thermodynamical limit, is derived in \cite{Eldad-Ivan}:
\begin{equation}
\label{detAK}
\caA_\ww= \det \(  1 -  K \),
\end{equation}
where the $N\times N$ matrix $K$ has matrix elements
\be \la{defKmatr} K_{jk} &=&\frac{ Q_j}{w_{j} -w_k+ \ii } \label{DefK}
\qquad\qquad (j,k=1,\dots, N)\, , \ee
and the weights $Q_j$ are obtained as the residues of the same
function at the roots $w_j$: \be Q_j &\equiv& \underset{z\to w_j}{
\text{Res}} \CQ(z), \quad \CQ (z) \equiv f(z)\ {Q_\ww(z+\ii)\over
Q_\ww(z)}
.
 \label{EjDef}
\ee
  Here $Q_\ww$ is the Baxter polynomial for the set $\ww$, c.f.
  \re{defBaxP}.  The determinant  formula \re{detAK} has the advantage
  that it exponentiates in a simple way:
\begin{eqnarray}
\la{logDet} \log\caA_{\ww} [f] =- \sum_{n=1}^\infty\ {1\over n}
\sum_{j_1, \dots, j_n=1}^N { Q_{j_1}\over w_{j_1}-w_{j_2}+\ii} \ {
Q_{j_2}\over w_{j_1}-w_{j_3}+\ii} \ \cdots \ { Q_{j_n}\over
w_{j_n}-w_{j_1}+\ii}.
\end{eqnarray}
The identity  \re{detAK} is the basis for the
field-theoretical approach to the computation of the scalar product
in the thermodynamical limit.

  \section{Field theory of the inner product}
  \la{FieldTheory}

  \subsection{The $\caA$-functional in terms of free fermions}

This determinant on the rhs of  \re{detAK}
can be expressed as a Fock-space expectation value for a Neveu-Schwarz
chiral fermion living in the rapidity  plane with two-point function
\be \la{opepsia} \bra \psi(z) \psi^*(u)\ket =  \bra
\psi^*(z) \psi(u)\ket = {1\over z-u}\, .
 \ee
 Representing  the matrix $K$ in \re{detAK} as
 \be
 K_{jk}= \bra \psi^*(w_j+\ii) \psi(w_k)
 \ket
 \ee
 it is easy to see that the $\caA$-functional is given by the expectation value
 \be \la{fermionrepD} \caA_\ww[f] = \bra \exp \( \sum_{j=1}^N Q_j \,
 \psi^*(w_j ) \psi(w_j+ \ii)\)\ket .  \ee

In order to take the large $N$ limit, we will need reformulate the
problem entirely in terms of the meromorphic function $\CQ(z)$.  The
discrete sum of fermion bilinears in the exponent on the rhs of
\re{fermionrepD} can be written as a contour integral using the fact
that the quantities $Q_j$, defined by \re{EjDef}, are residues of the same function
$\CQ(z)$ at $z=w_j$.  As a consequance, the Fock space representation
\re{fermionrepD} takes the form
 \be \la{fermionrepC} \caA_{\ww}[f]  = \bra \exp \( \oint_{\CC_\ww} {d
 z\over 2\pi i} \CQ (z) \, \psi^* (z ) \psi(z+\ii)\)\ket\, , \ee
where the contour $\CC_\ww$ encircles the points $\ww$ and leaves
outside all other singularities of $\CQ$,
as shown in Fig. \ref{fig:Contour}.  Expanding the exponent and
performing the gaussian contractions, one writes the $\caA$-functional
in the form of a Fredholm determinant
 \be
 \begin{aligned}
   \la{AGKPF} \caA_{\ww}[f]&= \sum_{n=0}^\infty {(-1)^n\over n!}
  \oint\limits _{\CC_\ww^{\times n}}  \prod_{j=1}^n {dz_j \, \CQ (z_j)\over 2\pi i} \
   \det_{j,k=1}^n {1\over z_j-z_k + \ii } .
   \end{aligned}
 \ee
Since the function $\CQ $ has exactly $N$ poles inside the contour
$\CC_\ww$, only the first $N$ terms of the series are non-zero.  The
series exponentiates to
 \be\la{expansionFred1} \log \caA_\ww[f]&=& - \sum _{n=1}^\infty\
 {1\over n} \oint\limits_{\CC_\ww^{\times n}} {dz_1\dots dz_n \over (2\pi
 i)^n} { \CQ (z_1)\over z_1-z_2+ \ii }\dots {\CQ (z_n)\over z_n-z_1+
 \ii }.  \ee
This is the vacuum energy energy of the fermionic theory, given by the sum of
all vacuum loops.  The factor $(-1)$ comes from  the Fermi statistics
and the factor $1/n$ accounts for the cyclic symmetry of the loops.
The series \re{expansionFred1} can be of course obtained directly from
\re{logDet}.

\subsection{Bosonic theory and Coulomb gas}

Alternatively, one can express the $\caA$-function in term of a chiral
boson $\phi(x)$ with two-point function
\be \bra \phi(z) \phi(u)\ket = \log(z-u).  \ee
After bosonization $\psi(z) \to e^{\phi(z)}$ and $\psi^*(z) \to
e^{-\phi(z)}$, where we assumed that the exponents of the gaussian
field are normally ordered, the fermion bilinear $\psi^*(z) \psi(z+
\ii)$ becomes, up to a numerical factor, a chiral vertex operator of
zero charge
\be \Vii  (z) \equiv e^{\phi(z+\ii) - \phi(z)}.  \ee
The coefficient is obtained from the OPE
\be \la{bosonicOPE} e^{-\phi(z)} \, e^{\phi(u)} \sim \ {1\over z-u} \
e^{\phi(u)-\phi(z)} \ee
with $u= z+\ii$:
\be
\la{defAvert}
\psi^*(z) \psi(z+ \ii)\quad \ \to\ \quad
  e^{-\phi(z)}e^{\phi(z+\ii) }
= - {1\over \ii}\,  \Vii  (z)  .
\ee
The bosonized form of the operator representation \re{fermionrepC} is
therefore
 \be
 \la{bosonrepD}
 \caA_\ww[f]
 = \bra \exp \( - {1\over i\ve} \oint _{\CC_\ww}{dz\over 2\pi i} \
 \CQ(z) \, \Vii  (z) \)\ket , \ee
where $|0\rangle$ is the bosonic vacuum state with zero charge.
Expanding the exponential and applying the OPE \re{bosonicOPE} one
writes the expectation value as the grand-canonical Coulomb-gas
partition function
  \begin{eqnarray}
  \la{expFDbos}
  \begin{aligned}
  \caA_{\ww}[f]&=
    \sum_{n=0}^N  {(-1)^n\over n!}
    \prod_{j=1}^n \oint_{\CC_\ww} {dz_j \over 2\pi i} \
   \ {\CQ  (z_j) \over \ii}\ \ \prod_{j<k}^n { (z_{j}-z_k)^2  \over
    (z_{j}-z_k)^2-  \ii   ^2 }
 \, .\end{aligned}
  \end{eqnarray}
 After applying the Cauchy identity, we get back the Fredholm
 determinant \re{AGKPF}.

\subsection{The thermodynamical limit}
\label{section:limit}

  Although the roots $\ww = \{w_1,\dots, w_N\}$ are off-shell,
  typically they can be divided into two or three on-shell subsets
  $\ww^{(k)}$, each representing a lowest energy solution of the Bethe
  equations for given (large) magnon number $N^{(k)}$.
  The Bethe roots for such solution are organised in one of several
  arrays with spacing  $\sim \ve$, called macroscopic Bethe strings,
  and the distribution of
the roots along these arrays is approximated by continuous densities
on a collection of contours in the complex rapidity plane
  \cite{PhysRevLett.74.816,PhysRevLett.85.2813,Beisert:2003xu,Kazakov:2004qf}.

We choose
an $N$-dependent normalisation of the rapidity such that $\ve\sim
1/N$.  Then the typical size of the contours and the densities
remains finite in the limit $\ve\to 0$.
 
 In order to compute the $\caA$-functional in the large $N$ limit,
 we will follow the method developed on
  \cite{Eldad-Ivan} and based on the field-theoretical
 formulation of the problem, eq.  \re{bosonrepD}.
 The method involves a coarse-graining procedure, as does the original
 computation of the quantity $\CA$, carried out in \cite{GSV}.

Let us mention that there is a close analogy between the above
 semiclassical analysis and the computation of the instanton partition
 functions of four-dimensional $\CN=2$ supersymmetric gauge theories
 in the so-called $\Omega$-background, characterised by two
 deformation parameters, $\ve_1$ and $\ve_2$ \cite{Moore:1998et,G.Moore:2000aa}, in the Nekrasov-Shatashvili limit $\ve_2\to 0$
 \cite{Nekrasov:2009aa}.  In this limit the result is expressed in
 terms of the solution of a non-linear integral equation.  The
 derivation, outlined in \cite{Nekrasov:2009aa} and explained in great
 detail in the recent papers \cite{Mayer-MY, JEB-Mayer}, is based on
 the iterated Mayer expansion for a one-dimensional non-ideal gas.
 Our method is a field-theoretical alternative of the the Mayer
 expansion of the gas of dipole charges created by the exponential
 operators $\CV_n$.  In our problem the saddle-point of the action
 \re{defYY} also lead to a non-linear integral equation, but the
 non-linearity disappears when $\ve\to 0$.

\begin{figure}
         \centering
     \begin{minipage}[t]{0.6\linewidth}
            \centering
            \includegraphics[width=7.2 cm]{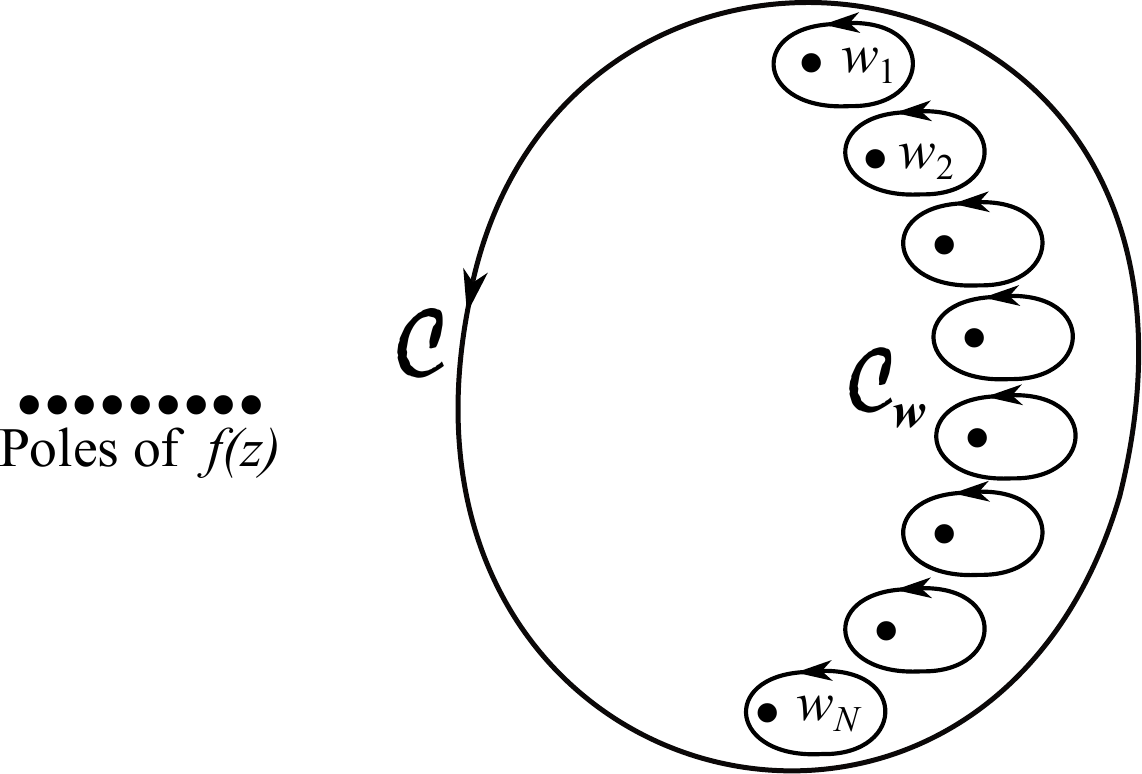}
  \caption{ \small Schematic representation of the
    contour $\CC_\ww$ and the deformed contour $\CC$.  }
  \label{fig:Contour}
         \end{minipage}%
           \end{figure}

Of crucial relevance to our approach is the possibility to deform
 the contour of integration.
 In order to take advantage of the contour-integral representation,
 the original integration contour $\CC_\ww$ surrounding the poles
 $\ww$ of the integrand, should be deformed to a contour $\CC$ which
 remains at finite distance from the singularities of the function
 $\CQ$ when $\ve\to 0$, as shown in Fig. \ref{fig:Contour}.
   Along the contour $\CC$ the function $\CQ(z)$
changes slowly at distances $\sim\ve$.  In all nontrivial
 applications the weight function $\CQ$ has a dditional poles, which
 are those of the function $f$.  The contour $\CC$ separates the roots
 $\ww$ from the poles of $f$.

     \subsection{ Coarse-graining}

   We would like to compute the $\ve$-expansion of the expectation
   value \re{bosonrepD}, with $\CC_\ww$ replaced by $\CC$.  This is a
   semi-classical expansion with Planck constant $\hbar=\ve$.  As any
   semi-classical expansion, the perturbative expansion in $\ve$ is an
   asymptotic expansion.  Our strategy is to introduce a cutoff $\L$,
   such that
  \be
  \ve\ll\L\ll  N\ve\qquad (N\ve\sim 1),
  \ee
  integrate the ultra-violet (fast-scale) part of the theory in order to
  obtain an effective infrared (slow-scale) theory.  The splitting of the
  bosonic field into slow and fast pieces into slow and fast pieces is
  possible only in the thermodynamical limit $\ve\to 0$.  In this
  limit the dependence on $\L$ enters through exponenttially small
  non-perturbative terms and the perturbative expansion in $\ve$ does
  not depend on $\L$.

 We thus cut the contour $\CC$ into segments of length $\L$ and
 compute the effective action for the slow piece as the sum of the
 connected $n$-point correlators (cumulants) of the vertex operator
 $\Vii $.  The $n$-th cumulant $\Xi_n(z)$ is obtained by integrating the
 OPE of a product of $n$ vertex operators
    \be \la{OPEA} \Vii(z_1)\dots \Vii(z_n)= \prod _{j<k} {(z_j-z_k)^2
    \over (z_j-z_k)^2+\ve^2} :\Vii(z_1)\dots \Vii(z_n): \ee
along a segment of the contour $\CC$ of size $\L$, containing the
point $z$.  Since we want to evaluate the effect of the short-distance
interaction due to the poles, we can assume that the rest of the
integrand is analytic everywhere.  Then the integration can be
performed by residues using the Cauchy identity.  This computation has
been done previously in \cite{G.Moore:2000aa} in a different context.
The easiest way to
compute the integral is to fix $z_1=z$ and integrate with respect to
$z_2, \dots, z_n$.  We expand the numerical factor in \re{OPEA} as a
sum over permutations.  The $(n-1)!$ permutations representing maximal
cycles of length $n$ give identical contributions to the residue.  For
the rest of the permutations the contour integral vanishes.  We find
($z_{jk} \equiv z_j-z_k$)
    \be
\la{MNScomputation} \Xi_n&=& \oint { \Vii(z_1)\dots \Vii(z_n)\over
(-\ii)^n \ n!} \prod_{k=2}^n {dz_k\over 2\pi i} \no \\
 &\sim & { (n-1)!\over n!}
\oint
{ \prod_{k=2}^n {dz_k\over 2\pi i}\ : \Vii (z_1)\cdots \Vii (z_n):
\over (\ii-z_{12} ) \dots(\ii-z_{n-1,n})(\ii- z_{n,1})}
\no \\
&=&-
{1\over n^2\ii} \ \CV_{n}(z)\, ,
\ee
where
 \be \la{nvertex} \CV_{n}(z) \equiv\ :\Vii (z) \Vii (z+\ii)\dots \Vii (z+
 n\ii): \ = e^ { \phi(z+ n\ii)-\phi(z) }\, .  \ee

The interaction potential of the effective coarse-grained theory
therefore contains, besides the original vertex operator $\Vii \equiv \CV_1$, all composite vertex operators $\CV_{n}$ with $n\lesssim \L$.
If one repeats the computation \re{MNScomputation} with the weights
$\CQ$, one obtains for the $n$-th cumulant
\be \Xi_n(z)&=&- {1\over \ii}\, {\CQ_{n}(z)\, \CV_{n}(z)\over
n^2}\, , \qquad \CQ_{n}(z) = \CQ(z) \CQ(z+i \ve)\dots \CQ(z+ i n\ve)
.
\ee
\be \Xi_n(z)&=&- {1\over \ii}\, {\CQ_{n}(z)\, \CV_{n}(z)\over
n^2}\, , \qquad \CQ_{n}(z) = \CQ(z) \CQ(z+i \ve)\dots \CQ(z+ i n\ve)
=e^{-\Phi(x)+\Phi(x+n\ii)} \,
.
\ee
As the spacing $n\ve$ should be smaller than the cut-off length $\L$,
from the perspective of the effective infrared theory all these
particles are point-like.  We thus obtained that in the semi-classical
limit the $\caA$-functional is given, up to non-perturbative terms, by
the expectation value
 \be
 \la{effectiveFT}
 \mathscr{A}_{\uu,\zz} \approx
 \< \exp \({1\over\ve } \sum_{n=1}^ {\L/\ve} {1\over n^2}
  \oint_{\CC} {d z\over 2\pi } \,
 \CQ_{n} (z)  \,  \CV_{n}(z)\)\> .
 \ee

The effective potential can be given a nice operator form, which will
be used to extract the perturbative series in $\ve$.  For that it is
convenient to represent the function $f(z)$ as the ratio
\be
f(z) = {g(z)\over g(z+\ii )}=
g(z)^{-1} \ID\,  g(z)\, ,
\ee
where we introduced  the shift operator
\be
\ID\equiv e^{\ii\p}\, .
\ee
 Then the weight factor $\CQ_n$  takes the form
 \be
 \CQ_n =e^{-\Phi}\, \ID^n\,  e^{\Phi},
 \qquad \Phi(z) = Q_\ww(z)/g(z)\, ,
 \ee
and the series in the exponent in \re{effectiveFT} can be summed up to
 \be
\la{bosonicAC1}
\begin{aligned}
  \mathscr{A}_{\ww} [f] 
  &= \< \exp\( {1\over \ve} \oint_{\CC} {dz\over
  2\pi } \ :e^{ -\Phi(z)-\phi(z) } \ \Li( \ID )\ e^{ \Phi(z)
  +\phi(z)}: \) \> \, ,
\end{aligned}
\ee
with the operator $\Li(\ID)$ given by the dilogarithmic series
\be
\Li(\ID)=\sum_{n=1}^\infty {\ID^n\over n^2}.
\ee
Here we extended the sum over $n$ to infinity, which which can be done
with exponential accuracy.  The function $\Phi(z)$, which we will
refer to as ``classical potential'', plays the role of classical
expectation value for the bosonic field $\phi$.

If we specify to the case of the (inhomogeneous,
twisted) spin chain, considered in \cite{Eldad-Ivan}, then $f= \k\,
d/a$ with $a,d$ given by \re{adinhom}.  In this case the classical
potential is
 \be \Phi(z) =\log Q_\ww(z)- \log Q_\thth(z- \ii/2).  \ee

 \bigskip

 {\it Remark.} Going back to the fermion representation, we write
 the result as a Fredholm determinant with different Fredholm kernel,
 \be \la{fermionrepCdsc} \caA_{\ww} [f] \approx \bra \exp
 \(\oint_{\CC} {d z\over 2\pi i} e^{-\Phi(z) }\psi^* (z )\, \log(1-
 \ID)\, \psi(z) e^{\Phi(z)}\)\ket = \Det (1-\hat \CK), \ee
where the Fredholm  operator $\hat\CK$ acts in the space of functions
analytic in the vicinity of the contour $\CC$:
\be
\hat\CK\xi(z) =\oint_\CC {du\over 2\pi i }  \hat \CK(z, u)\xi(u)
, \qquad
  \hat \CK(z, u)= \sum_{n=1}^\infty
{e^{-\Phi(z) +\Phi(z+\ii n)}\over z-u+\ii n} .
\ee
The expression in terms of a Fredholm determinant can be obtained
directly by performing the cumulant expansion for the expression of
the $\caA$-functional as a product of shift operators \cite{SL}
\be
\begin{split}
\caA[f]&= {1\over \Psi_\ww[g]} \prod_{j=1}^N (1- e^{\ii \p/\p w_j})
\prod_{j=1}^N \Psi_{\ww}[g], \\
 \Psi_\ww[g] &=
{\prod_{j<k} (w_j-w_k)\over \prod_{j=1}^N g(w_j)},
\quad f(z) = {g(z)\over g(z+\ii)}.
\end{split}
\ee

       \subsection{ The first two orders of the semi-classical
       expansion}

 The effective IR theory is compatible with the semi-classical
 expansion being of the form
 \be
 \log \caA_{\ww} = {F_0 \over \ve} + F_1 + \ve  F_2 + \dots
 + \CO(e^{-\L/\ve}).
 \ee
Below we develop a diagram technique for computing the coefficients in
the expansion.  First we notice that the $\ve$-expansion of the
effective interaction in \re{bosonicAC1} depends on the field $\phi$
through the derivatives $\p\phi, \p^2\phi$, etc.  We therefore
consider the first derivative derivative $\p\phi$ as an independent
field
 \be \vp(z) \equiv - \p\phi(z) \ee
 with two-point function
 \be
 \la{tpf}
 G(z,u) = \p_z\p_u\log(z-u) = {1\over (z-u)^2}.
 \ee
In order to derive the diagram technique, we formulate the expectation
value \re{bosonicAC1} as a path integral for the $(0+1)$-dimensional
field $\vp(x)$ defined on the contour $\CC$.  The two-point function
\re{tpf} can be imposed in the standard way by introducing a second
field $\rho(x)$ linearly coupled to $\vp$.  The path integral reads
\be
\mathscr{A}_{\ww} [f] &=&
 \int [D\vp\, D\rho]\ e^{- \CY[\vp, \rho]}\, ,
\la{pathint}
 \ee
 with action functional
\be \la{defYY} \CY[\vp,\rho] = -\hf \int\limits_{\CC\times \CC} {dz}
{du}\ {\rho(z) \rho(u)\over (z-u)^2} + \oint_\CC dx\, \rho(z) \vp(z) +
\oint_\CC {dz\over 2\pi} \ {W}(\vp, \vp',\dots)\, .  \ee

The dependence on $\ve$ is through the potential $W$, obtained by
expanding the exponent in \re{bosonicAC1}:
\be \la{defWW} W(\vp, \vp',\dots) & =& - {1\over \ve} \
e^{-\Phi(x)-\phi(x)} \ \Li(\ID )\ e^{\Phi(x)+\phi(x)} \no \\
    &=& - {1\over \ve} \ \Li ( \CQ ) +i \log(1- \CQ) \vp - {\ve \over
    1-\CQ} (\vp^2+\vp')+ \CO(\ve^2).  \ee
The potential contains a constant term, which gives the leading
contribution to the free energy, a tadpole of order $1$ and higher
vertices that disappears in the limit $\ve\to 0$.
The Feynman rules for the effective action $\CY[\vp, \rho]$ are such
that each given order in $\ve$ is obtained as a sum of finite number
of Feynman graphs.  For the first two orders one obtains
\be \la{F-zero} F_0&=& \oint \limits_{\CC} {dx\over 2\pi } \ \Li[
\CQ(x)]\, , \\
\la{F-one} F_1&=& - \hf \oint\limits_{ \CC\times\CC} {dx\, du \over
(2\pi )^2}\ { \log\[ 1- \CQ(x) \] \ \log\[ 1-\CQ(u) \]\over (x-u)^2} .
\ee
 where the double integral is understood as a principal value.  The
 actual choice of the contour $\CC$ is a subtle issue and depends on
 the analytic properties of the function $\CQ(x)$.  The contour shold
 be placed in such away that it does not cross the cuts of the
 integrand.

 \begin{figure}
 \begin{center}
     \begin{minipage}[t]{0.8\linewidth}
            \centering
            \includegraphics[width=8.5 cm]{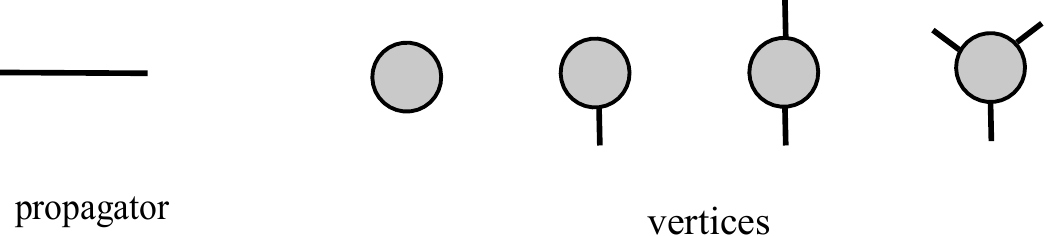}
\caption{ \small Feynman rules for the effective field theory }
  \label{fig:Feynman}
         \end{minipage}%
\end{center}
\end{figure}

 \begin{figure}
 \begin{center}
     \vskip 1cm
     \begin{minipage}[t]{0.8\linewidth}
            \centering
            \includegraphics[width=4.8 cm]{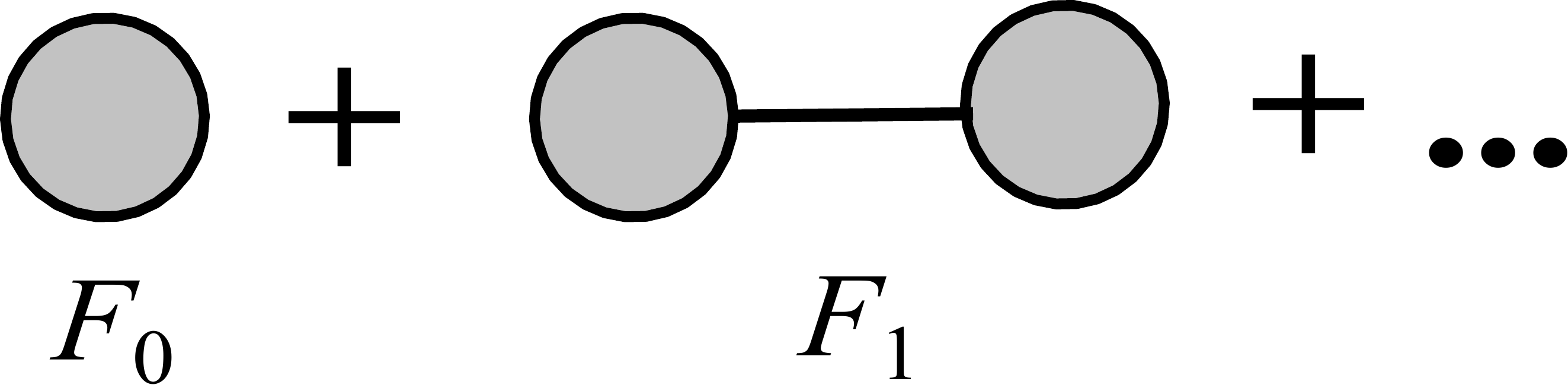}
\caption{ \small The leading and the subleading orders of the vacuum energy }
  \label{fig:Vacuumenergy}
         \end{minipage}%
\end{center}\end{figure}

Returning to the scalar product and ignoring the trivial factors in
\re{defSuv0}, we find that the first two coefficients of the
semi-classical expansion are given by eqs.  \re{F-zero} and \re{F-one}
with
\be \CQ=  e^{i p_\uu + i p_\vv} .
\ee
%
%

     \section{Discussion}

 In these notes we reviewed the field-theoretical approach to the computation
 of  scalar
 products of on-shell/off-shell Bethe vectors in the generalised model
 with $SU(2)$ rational $R$-matrix, which leads to a systematic procedure
 for computing the
 semi-classical expansion.  The results reported here represent a
 slight generalisation if those already reported in \cite{SL, sz,
 Eldad-Ivan}.
 We hope that the field-theoretical method  could be used to compute
scalar products in integrable models associated with higher rank  groups,
using the fact that the
 the integrands in the multiple contour integrals of in
  \cite{2005TMP...145.1373P,
2009AnHP...10..513F,2010SIGMA...6..094B,2013arXiv1306.0552W}
is expressed as  products of $\caA$-functionals.

The problem considered here is formally similar to the problem of computing the
instanton partition functions in $\CN=1$ and $\CN=2$ SYM  \cite{Moore:1998et,G.Moore:2000aa,Nekrasov:2009aa}.
As a matter of fact, the scalar product in the form \re{expFDbos} is the grand-canonical
 version of the partition function of the $\CN=1$ SUSY in four dimensions,
 which was studied in a different large $N$ limit in \cite{Kazakov:1998ji}.

 Our main motivation was the computation of the three-point function
 of heavy operators in $\CN=4$ four-dimensional SYM. Such operators
 are dual to classical strings in $AdS_5\times S^5$ and can be
 compared with certain limit of the string-theory results.  For a
 special class of three-point functions, the semi-classical expansion
 is readily obtained from that of the scalar product.  The leading
 term $F_0$ should be obtained on the string theory side as the
 classical action of a minimal world sheet with three prescribed
 singularities.  The comparison with the recent computation in
 \cite{KKnew} looks very encouraging. 
We expect that the meaning of the
 subleading term on the string theory side is that it takes account of
 the gaussian fluctuations around the minimal world sheet.
In this context it would be interesting  to obtain the subleading order  of the
 heavy-heavy-light correlation function in the $su(2)$ sector  in string theory \cite{Zarembo:2010ab, Costa:2010rz, EGSV:Tailoring2}.
 In the near-plane-wave limit the subleading order  was
 obtained in \cite{Klose:2013aza}.

  \section*{Acknowledgments}
  The author thanks E. Bettelheim, N. Gromov, T. Mc Loughlin
  and S. Shatashvili and  for
  valuable discussions.
   This work has been supported by European Programme IRSES UNIFY (Grant No 269217).

{ \footnotesize


\providecommand{\href}[2]{#2}\begingroup\raggedright\endgroup
  }

 \end{document}